\documentclass[aps,prl,twocolumn,showpacs,superscriptaddress]{revtex4-1}  
\usepackage{graphicx}  
\usepackage{dcolumn}   
\usepackage{bm}        
\usepackage{amssymb}   
\usepackage{epstopdf}
\usepackage{amsmath}
\usepackage{natbib}
\usepackage{multirow}
\usepackage{ulem}
\usepackage{color}
\usepackage{soul}

\begin{document}


\title{Magnetotransport and de Haas-van Alphen measurements in the type-II Weyl semimetal TaIrTe$_4$}

\author{Seunghyun Khim}
\email [E-mail: ]{Seunghyun.Khim@cpfs.mpg.de}
\affiliation{Leibniz-Institute for Solid State and Materials
    Research, IFW Dresden, Helmholtzstra{\ss}e 20, 01069 Dresden, Germany}
\affiliation{Max-Planck-Institut f\"{u}r Chemische  Physik fester Stoffe, N\"{o}thnitzer Strasse 40, 01187, Dresden}

\author{Klaus Koepernik}
\affiliation{Leibniz-Institute for Solid State and Materials Research, IFW Dresden, Helmholtzstra{\ss}e 20, 01069 Dresden, Germany}

\author{Dmitry V. Efremov}
\affiliation{Leibniz-Institute for Solid State and Materials Research, IFW Dresden, Helmholtzstra{\ss}e 20, 01069 Dresden, Germany}

\author{J. Klotz}
\affiliation{Hochfeld-Magnetlabor Dresden (HLD-EMFL), Helmholtz-Zentrum Dresden-Rossendorf, 01328 Dresden, Germany}
\affiliation{Department of Physics, TU Dresden, D-01062 Dresden, Germany}

\author{T. F\"{o}rster}
\affiliation{Hochfeld-Magnetlabor Dresden (HLD-EMFL), Helmholtz-Zentrum Dresden-Rossendorf, 01328 Dresden, Germany}
\affiliation{Department of Physics, TU Dresden, D-01062 Dresden, Germany}

\author{J. Wosnitza}
\affiliation{Hochfeld-Magnetlabor Dresden (HLD-EMFL), Helmholtz-Zentrum Dresden-Rossendorf, 01328 Dresden, Germany}
\affiliation{Department of Physics, TU Dresden, D-01062 Dresden, Germany}

\author{Mihai I. Sturza}
\affiliation{Leibniz-Institute for Solid State and Materials Research, IFW Dresden, Helmholtzstra{\ss}e 20, 01069 Dresden, Germany}

\author{Sabine Wurmehl}
\affiliation{Leibniz-Institute for Solid State and Materials Research, IFW Dresden, Helmholtzstra{\ss}e 20, 01069 Dresden, Germany}
\affiliation{Department of Physics, TU Dresden, D-01062 Dresden, Germany}

\author{Christian Hess}
\affiliation{Leibniz-Institute for Solid State and Materials Research, IFW Dresden, Helmholtzstra{\ss}e 20, 01069 Dresden, Germany}
\affiliation{Department of Physics, TU Dresden, D-01062 Dresden, Germany}

\author{Jeroen van den Brink}
\affiliation{Leibniz-Institute for Solid State and Materials Research, IFW Dresden, Helmholtzstra{\ss}e 20, 01069 Dresden, Germany}
\affiliation{Department of Physics, TU Dresden, D-01062 Dresden, Germany}
\affiliation{Department of Physics, Harvard University, Cambridge, Massachusetts 02138, USA}

\author{Bernd B\"{u}chner}
\affiliation{Leibniz-Institute for Solid State and Materials Research, IFW Dresden, Helmholtzstra{\ss}e 20, 01069 Dresden, Germany}
\affiliation{Department of Physics, TU Dresden, D-01062 Dresden, Germany}\date{\today}

\begin{abstract}
The layered ternary compound TaIrTe$_4$ has been predicted to be a type-II Weyl semimetal with only four Weyl points just above the Fermi energy. Performing magnetotransport measurements on this material we find that the resistivity does not saturate for fields up to 70 T and follows a $ \rho \sim  B^{1.5}$ dependence. Angular-dependent de Haas-van Alphen (dHvA) measurements reveal four distinct frequencies. Analyzing these magnetic quantum oscillations by use of density functional theory (DFT) calculations we establish that in TaIrTe$_4$ the Weyl points are located merely $\sim$ 40-50 meV above the chemical potential, suggesting that the chemical potential can be tuned into the four Weyl nodes by moderate chemistry or external pressure, maximizing their chiral effects on electronic and magnetotransport properties.
\end{abstract}

\pacs{}
\maketitle

A recent conceptual breakthrough in the theory and classification of metals is the discovery of Weyl semimetals \cite{Wan_2011,Balents_2011,Shekhar2015}. These semimetals have a topologically nontrivial electronic structure with fermionic Weyl quasiparticles -- massless chiral fermions that play as well a fundamental role in quantum field theory and high-energy physics \cite{Weyl1929}. A consequence is that in Weyl semimetals topologically protected surface states appear in the form of  Fermi lines that connect Weyl points (WPs) of opposite chirality, commonly referred to as Fermi arcs. 

Last year it was discovered that actually {\it two} types of Weyl fermions may exist in solids \cite{soluyanov2015}. Weyl semimetals of type-I have a point-like Fermi surface and consequently zero density of states at the energy of WPs~\cite{Burkov_2011,Xu_2011,Xu_2015,Yang_2015,Lv_2015,SYXu15,DFXu15,Belopolski15,Souma15,NXu15,Chang2015,Weng_2015,Huang_2015}. This is very different from Weyl semimetals of type-II~\cite{Volovik14,soluyanov2015}, which have thermodynamic density of states at the energy of Weyl nodes and acquire exotic Fermi surfaces: in type-II systems Weyl nodes appear at touching points between electron and hole pockets. The presence of these very peculiar states is predicted to strongly affect magnetotransport properties of a Weyl semimetal and causes the conduction of electric current only in certain directions in presence of a magnetic field~\cite{soluyanov2015,Zyuzin_2016,Ruan15}. In spite of the considerable progress made by theory, only a handful of type-II Weyl semimetals have been identified on the basis of electronic band-structure calculations: WTe$_2$, MoTe$_2$, Ta$_3$S$_2$, YbMnBi$_2$ and, very recently, TaIrTe$_4$ \cite{soluyanov2015,wang15,Sun2015,Chang2015,Borisenko2015,Koepernik2016}.

Of interest is in particular the orthorhombic ternary compound TaIrTe$_4$ as it combines structural simplicity with topological WPs: TaIrTe$_4$ is a structurally layered material which hosts just four type-II WPs, the minimal number of WPs a system with time-reversal invariance can host~\cite{wang15}. Moreover, the WPs are well separated from each other in momentum space. Such a large momentum-space separation promises a strong impact of the Weyl fermions on the transport properties. Indeed, we present in this Letter magnetotransport and magnetic quantum oscillations studies of TaIrTe$_4$ that evidence a non-saturating magnetoresistance signaling the presence of Weyl nodes. Analyzing de Haas-van Alphen (dHvA) oscillations by use of density functional theory (DFT) calculations we establish that in our TaIrTe$_4$ crystals the WPs are located $\sim$ 40-50 meV above the chemical potential, suggesting that the WPs might even match the Fermi energy by a slight tuning by chemistry or external pressure, which will maximize the effect of their chirality on the electronic properties, in particular on the magnetotransport.

\begin{figure}
\centering
\includegraphics[width=80mm]{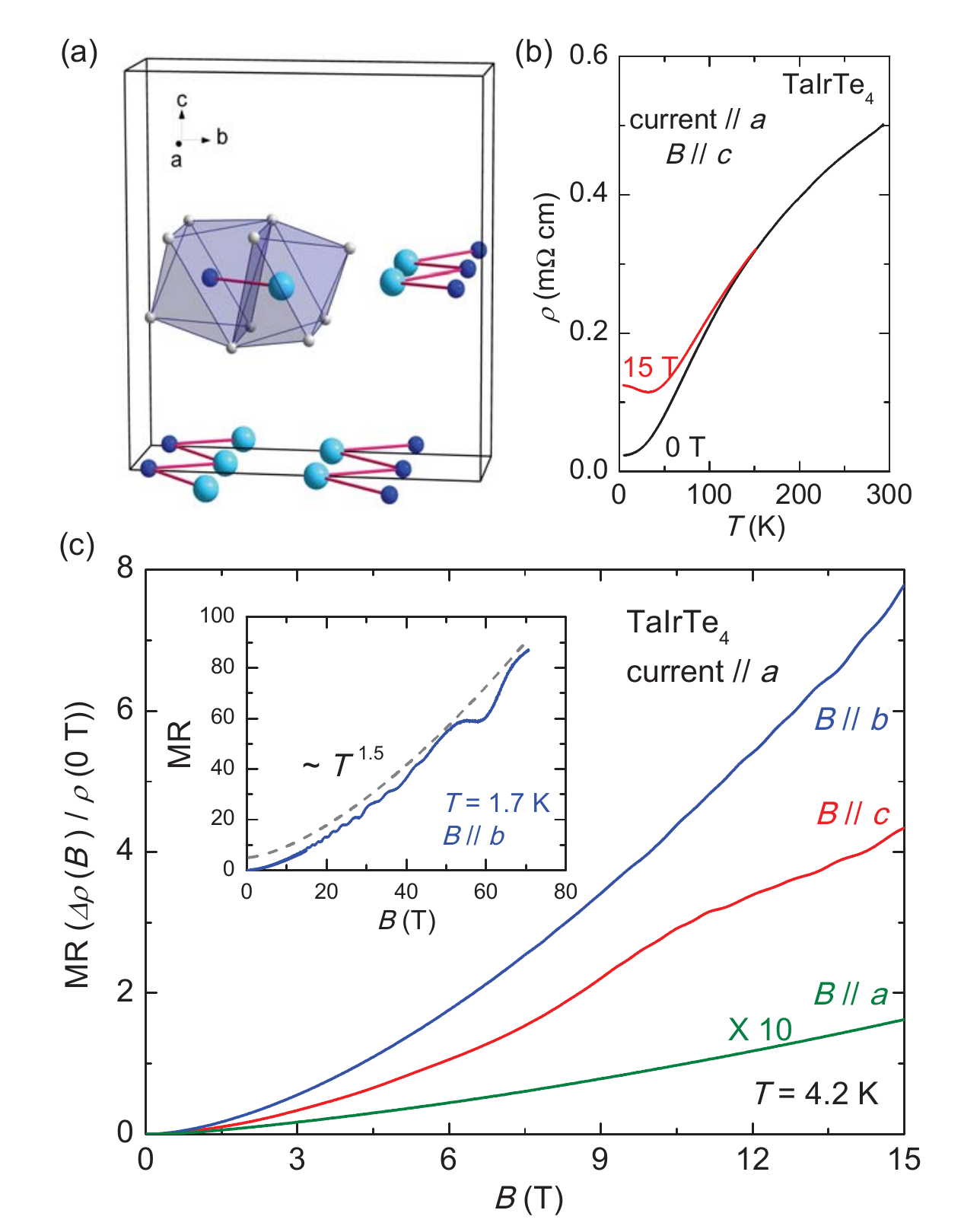}
\caption{\label{structure} (a) Crystal structure of TaIrTe$_4$. Blue and cyan balls represent Ta and Ir atoms, respectively. Te atoms (gray balls) are only drawn in the polyhedron for clarity. (b) Temperature dependence of the $a$ axis resistivity measured in {\it B} =  0 and 15 T applied along the $c$ direction. (c) Field-dependent magnetoresistance with fields along the three crystallographic directions. The MR data for $\mathbf{B}{\parallel}$$\mathbf{a}$ is magnified by 10 times for clarity. The inset shows the MR for $\mathbf{B}{\parallel}\mathbf{b}$ up to 70 T at $T$ = 1.7 K.} 
\end{figure}

Single crystalline TaIrTe$_4$ was grown from excess Te flux. Details of the crystal growth are described in the Supplemental Material \cite{supp}. We used three single crystals from the same batch for each of the resistivity, Hall-effect, and magnetic-torque measurements. The observed quantum-oscillation frequencies in the three samples closely agree with each other, evidencing that they have the same electronic structure and chemical potential. The crystals have a needle-like shape that is due to their crystal structure consisting of zigzag chains of alternating Ta-Ir connections along the $a$ direction, see Fig. \ref{structure}(a). These chains hybridize with each other along the $b$ direction to form a conducting $ab$ plane. This compound shares the space group (Pmn2$_1$) with WTe$_2$ while its unit cell is doubled along the $b$ direction. The resistivity and Hall measurements were done by using the conventional four- and five-probe methods in a superconducting magnet up to 15 T. Resistivity in pulsed fields up to 70 T was measured in the Hochfeld-Magnetlabor Dresden at Helmholtz-Zentrum Dresden-Rossendorf.

The resistivity measured along the $a$ direction [Fig. \ref{structure}(b)] shows metallic behavior in zero magnetic field with a residual resistivity ratio RRR = $\rho$(292 K)/$\rho$(4.2 K) = 21.5. In a magnetic field $\mathbf{B}{\parallel}\mathbf{c}$ the resistivity strongly increases and even qualitatively changes its temperature ($T$) dependence below about 30 K. Such a pronounced positive magnetoresistance is present for all three orientations of the magnetic field as displayed in Fig. \ref{structure}(c) where we plot the field dependence of the MR ratio, [$\rho$($B$)-$\rho$(0 T)]/$\rho$(0 T). While the qualitative behavior of the MR is similar, the absolute value is strongly anisotropic with the smallest increase for $\mathbf{B}{\parallel}\mathbf{a}$. The MR ratio for $\mathbf{B}{\parallel}\mathbf{b}$ (780 \%) at 15 T is nearly two times larger than that for $\mathbf{B}{\parallel}\mathbf{c}$ (420 \%).

The theoretically predicted chiral anomaly in Weyl semimetals manifests itself in a negative longitudinal MR, due to the Adler-Bell-Jackiw chiral anomaly \cite{Nielsen1983,Son2013} which has been observed in type-I Weyl semimetals such as NbP \cite{Zhang2015,Klotz2016}, TaAs \cite{Huang2015}, and TaP \cite{Zhang2015,Arnold2015,Hu2016}. The chiral anomaly in type-II Weyl semimetals is more subtle, because it is restricted to certain directions \cite{soluyanov2015}. Experimentally, our data for TaIrTe$_4$ as well as earlier results for WTe$_2$ show a positive longitudinal MR for $\mathbf{B}{\parallel}\mathbf{E}{\parallel}\mathbf{a}$ \cite{Ali2014,Rhodes2015}. 

Since there exist also conventional bands, it is very difficult to unambiguously separate the contribution from isolated Weyl points in transport measurements. Therefore, it is not possible to judge the presence or absence of the chiral anomaly in TaIrTe$_4$ from our present experimental data for the longitudinal MR. Anomalous behavior is, however, found for the transverse MR in TaIrTe$_4$: the resistivity increases in the magnetic field with $ \rho \propto B^{1.5}$ up to 70 T without saturation, see inset in Fig. \ref{structure}(c). It is remarkable that a very similar observation has been reported for WTe$_2$ \cite{Ali2014}. In a classical picture the large non-saturating transverse MR is due to the vicinity to a perfect balance between electron and hole carriers \cite{Ali2014}.
However, recently this observation has also been associated with the nontrivial chiral bands of Weyl semimetals \cite{Neupane2014,Liang2015}.

\begin{figure}
\centering
\includegraphics[width=80mm]{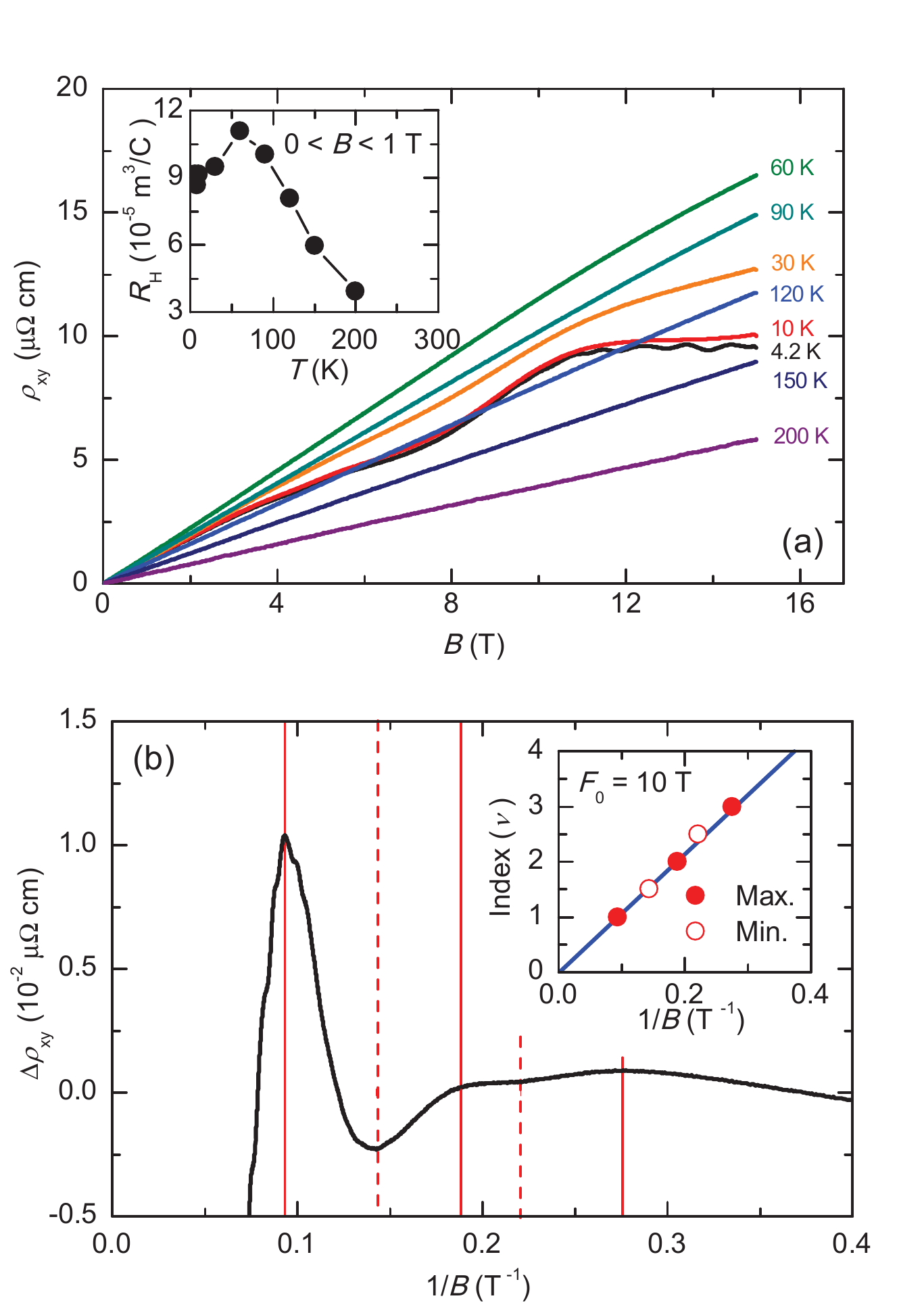}
\caption{\label{Hall} (a) Field-dependent Hall resistivity $\rho_{xy}$ at various fixed $T$. The inset shows the Hall coefficient determined below 1 T. (b) Derivative of $\rho_{xy}$ ($T$ = 4.2 K) versus 1/{\it B}. Vertical solid (dashed) lines mark maximum (minimum) positions on the oscillation feature. Assigned Landau level indices ($\nu$) as a function of 1/{\it B} are shown in the inset.}
\end{figure}

The Hall resistivity ($\rho_{xy}$) as a function of magnetic field for various temperatures is depicted in Fig. \ref{Hall}(a). The observed $\rho_{xy}$ is positive in the whole {\it T} interval, indicating dominating hole carriers. While $\rho_{xy}$ linearly increases with {\it B} at high $T$, a slowly oscillating feature appears below 30 K. From the derivative of the 4.2 K data, we identify a periodicity with 1/{\it B} and the corresponding frequency ($F$) is 10 T which is denoted as $F_0$ hereafter. This oscillation reaches the quantum limit above 10 T [inset of Fig. \ref{Hall}(b)]. To obtain the Hall coefficient ($R_H$ = ${\rho}_{xy}$/{\it B}), we extract the linear slope below 1 T, where the oscillatory feature is negligible. The strong $T$-dependent $R_H$ implies the presence of multiple charge-carrier channels. $R_H$ increases when lowering $T$, starting from 200 K, but decreases below $\sim$ 60 K as shown in the inset of Fig. \ref{Hall}(a). 

Qualitatively, such an anomalous $T$ dependence of $R_H$ is signaling bands with small hole and/or electron pockets consistent with the band structure of TaIrTe$_4$ discussed below. We mention that we did not observe noticeable changes in the resistivity as well as in Kohler plots from the MR data around 60 K (See the Supplemental Material \cite{supp}).

\begin{figure}
\centering
\includegraphics[width=80mm]{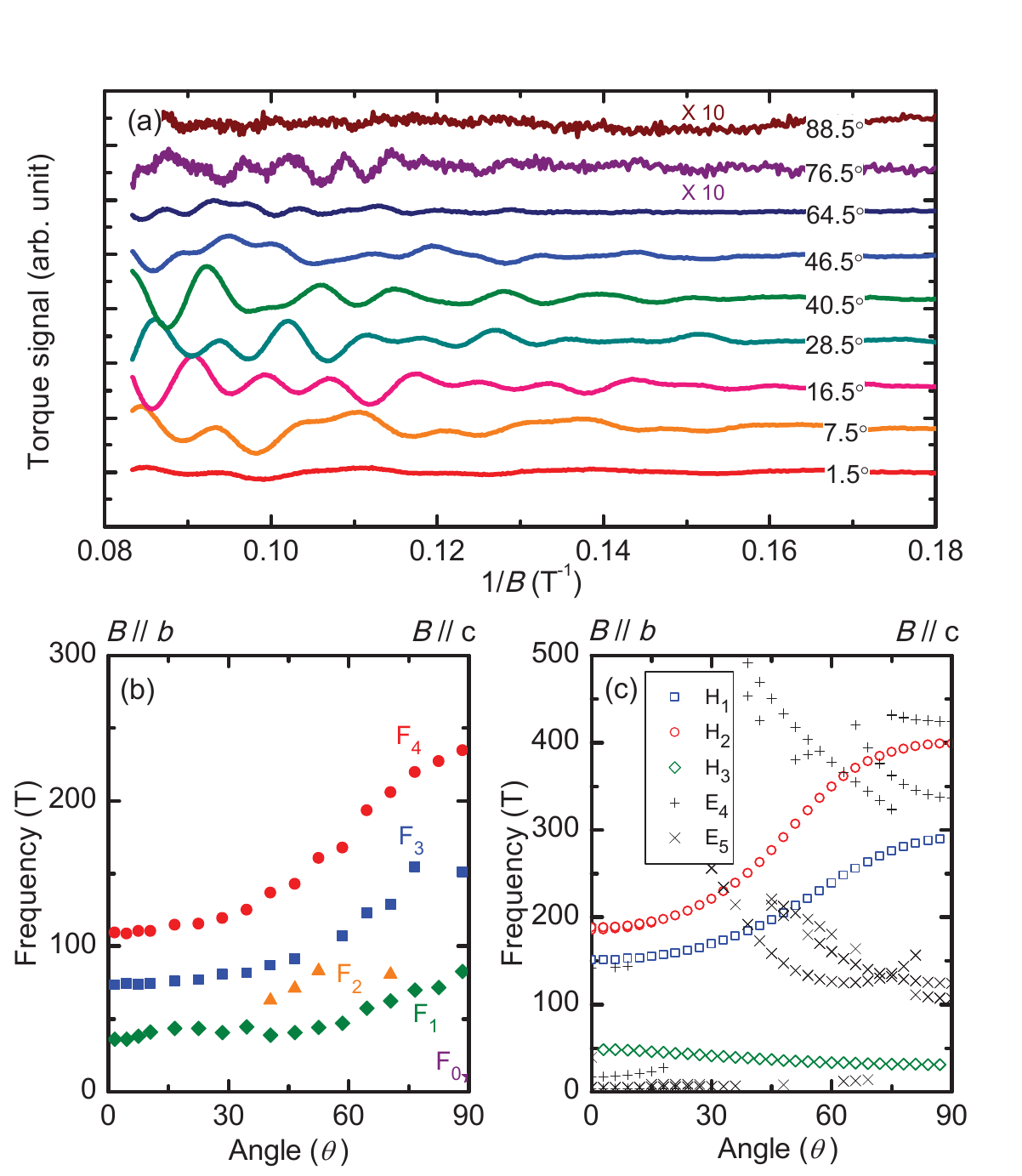}
\caption{\label{angle} (a) Oscillating torque signals vs 1/$B$ measured at various angles at $T$ = 1.5 K after subtracting a polynomial background. (b) Angular dependence of the dHvA frequencies (filled symbols). (c) Calculated angular dependence of dHvA frequencies ($E_F$ = 0).}
\end{figure}

We investigated the magnetic quantum oscillations in more detail in order to elucidate the electronic structure of TaIrTe$_4$. Magnetic-torque measurements up to 12 T were carried out by use of a capacitive 50 $\mu$m thick CuBe-foil cantilever placed in a superconducting magnet. We measured the torque dHvA signals at various angles at {\it T} = 1.5 K. The crystal was mounted in a configuration such that the applied field aligned along the $b$ axis for $\theta$ = $0^\circ$ and along the $c$ axis for $\theta$ = $90^\circ$. Figure \ref{angle}(a) shows the evolution in the torque signals. Varying from $\mathbf{B}{\parallel}\mathbf{b}$ to $\mathbf{B}{\parallel}\mathbf{c}$, the oscillating feature clearly changes, evidencing anisotropic three-dimensional Fermi-surface pockets.

Applying a fast Fourier transform (FFT) after subtracting a polynomial background from the magnetic torque signal, we obtained the corresponding frequency spectra. The observed frequencies are related to the extremal cross-sections of the Fermi surface ($A$) described by the Onsager relation, $F_i$ = $A_{i}{\hbar}/(2{\pi}e)$, where $\hbar$ is the reduced Planck constant and $e$ is the electron charge. Figure \ref{angle} (b) shows the angle-resolved $F$ plot. For $\mathbf{B}{\parallel}\mathbf{b}$, we identify frequencies of 36, 73.5, and 109 T which are denoted as $F_1$, $F_3$, and $F_4$, respectively. These frequencies smoothly grow as a field is tilted toward $\mathbf{B}{\parallel}\mathbf{c}$. We additionally detected $F_2$ for intermediate angles. The effective masses ($m^*$) were determined by analyzing $T$-dependent oscillation amplitudes at the fixed angle of ${\theta}$ = $46.5^{\circ}$. According to the Lifshitz-Kosevich formula, the oscillation amplitude is proportional to $X$/sinh($X$), where $X$ = ${\alpha}m^{*}T$/{\it B} and $\alpha$ = 2${\pi}k_{B}m_{e}$/(${\hbar}e$), with the Boltzmann constant $k_B$ and the free electron mass $m_e$. The resultant $m^{*}$/$m_e$ are 0.199 $\pm$ 0.003, 0.367 $\pm$ 0.005, 0.368 $\pm$ 0.006, and 0.370 $\pm$ 0.006 for $F_1$, $F_2$, $F_3$, and $F_4$, respectively (see the Supplemental Material for details \cite{supp}).

\begin{figure}
\centering
\includegraphics[width=85mm]{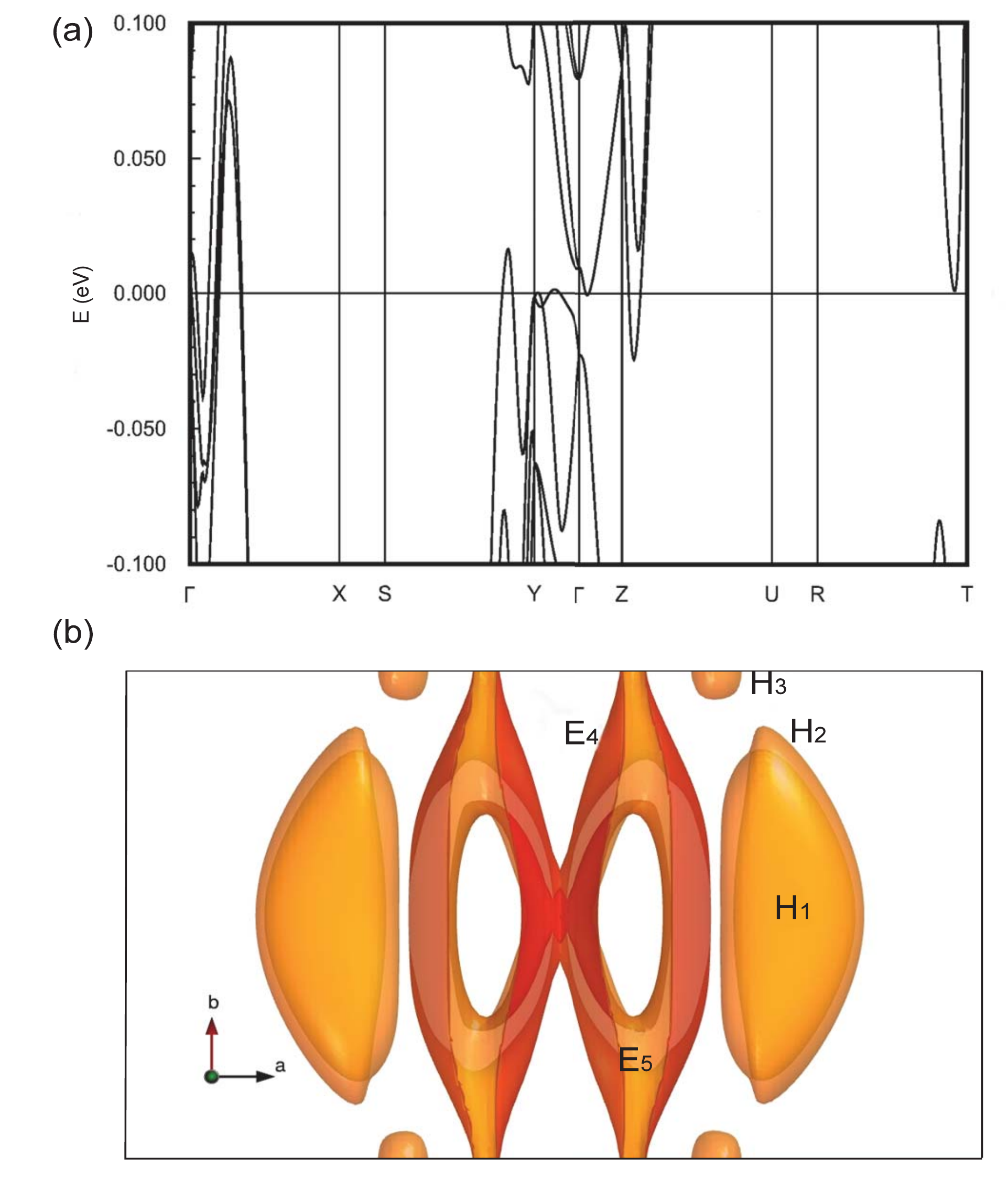}
\caption{\label{cal} (a) Calculated band structure of TaIrTe$_4$. (b) Fermi surfaces viewed along the -$k_c$ direction. The pockets H$_1$ and H$_2$ denote the inner and outer hole pockets. The pockets E$_4$ and E$_5$ are corrugated cylindrical electron pockets.}
\end{figure}

To interpret the data and establish their ramifications for the electronic structure of our crystals, we compare the measured frequencies with the results of DFT based electronic-structure calculations \cite{Koepernik2016}. Figure \ref{cal}(a) shows the electronic band structure in the vicinity of the Fermi level. The corresponding Fermi surface at the Fermi level is shown in Fig. \ref{cal}(b). Spin-orbit coupling splits the bands such that the hole pocket H$_1$ is contained in the hole pocket H$_2$, which also has a small disconnected part (H$_3$). The electron sheets showing a large anisotropy with open orbits along the $c$ direction, also split into the inner electron pocket E$_4$ and the outer electron pocket E$_5$.

Comparing theory and our experimental results it is straightforward to associate the frequencies $F_3$ and $F_4$ to extremal areas of the hole pockets H$_1$ and H$_2$, respectively. Apparently, the corresponding branches show very similar angular dependences for the experimental [Fig. \ref{angle}(b)] and calculated frequencies [Fig. \ref{angle}(c)]. 
The angular dependence of the frequency $F_1$ which is suggested by preliminary ARPES data to stem from the pocket H$_3$ is not properly reproduced in the calculations \cite{borisenko2016}.
The theoretically predicted signatures E$_4$ and E$_5$ are not observed in our experimental data. This discrepancy can be due to distinctive characters of these bands resulting in a weaker dHvA oscillation amplitude, for instance, a heavier effective mass or large scaattering rate.

Even if we restrict our discussion to the frequencies $F_3$ and $F_4$, where the assignment of the branches to the hole pockets H$_1$ and H$_2$ is straightforward, clear quantitative differences between theory and experiment are present. As apparent from Figs. 3(b) and 3(c), the experimental frequencies are smaller than the calculated ones. The most simple approach to account for this difference is assuming a rigid uniform shift of the chemical potential. Following such an approach, i.e., adjusting the frequencies $F_3$ and $F_4$ for $\mathbf{B}{\parallel}\mathbf{b}$, yields a chemical potential about 30 meV above the calculated Fermi level. Following the same procedure for $\mathbf{B}{\parallel}\mathbf{c}$ implies a further shift upwards by about 10-20 meV (See the Supplemental Material for details \cite{supp}). The limitations of this ``rigid-shift approach'' are obvious already from this anisotropy. More important, the deviation of the experimental $F_1$ oscillation from the theory is also not solved by rigid shifts of the calculated bands.

Despite these limitations the direction of the shift, which is justified for the two hole bands H$_1$ and H$_2$, is very interesting. Recent electronic-structure calculations have predicted type-II Weyl points in TaIrTe$_4$ located about 79 meV above the calculated Fermi energy \cite{Koepernik2016}. As the chemical potential in our crystals might be $\sim$ 30-40 meV above this Fermi level, the chemical potential would be just $\sim$ 40-50 meV below the energy of the Weyl nodes. 
Therefore, one might infer that the vicinity to the Weyl nodes is already important for the magnetotransport. Indeed, the anomalous behavior reported above, as for example the huge and strongly anisotropic magnetoresistance that does not saturate for fields up to 70 T, is supporting this qualitative conclusion. Thus, TaIrTe$_4$ is a very promising material for obtaining Weyl nodes very close to the Fermi level, for example by adjusting the synthesis conditions for the crystal growth or applying external pressure. It is also a very promising material for photoemission spectroscopy experiments aiming to resolve the Fermi arcs.
These experiments will benefit from the fact that the Weyl points are well separated: the length of the emerging Fermi arc is about 1/3 of the surface Brillouin zone. Moreover, the arcs are present at the natural (001) cleavage plane of TaIrTe$_4$ when its chemical potential is adjusted to the position of the Weyl nodes in the electronic structure.

\begin{acknowledgments}
We thank S. Borisenko, E. Hassinger, and J. Park for helpful discussions, and we acknowledge the support of the HLD at HZDR, member of the European Magnetic Field Laboratory (EMFL). This work was supported by the DFG through the Collaborative Research Center SFB 1143. JvdB acknowledges support from the Harvard-MIT CUA. 
\end{acknowledgments}

\bibliographystyle{apsrev}

\end{document}